\documentclass[12pt,a4paper]{article}
\usepackage[latin1] {inputenc}
\usepackage{amsmath}
\usepackage{amsfonts}
\usepackage{amssymb}
\usepackage[english]{babel}
\usepackage[dvips]{graphicx,color}
\title{Massless Fermions and the Instanton Dipole Liquid in Compact $QED_3$}
\author{C.~D.~Fosco\\ Centro At\'omico Bariloche and Instituto Balseiro\\
  Comisi\'on Nacional de Energ\'{\i}a At\'omica\\
  R8402AGP Bariloche, Argentina\\ \\
  L.~E.~Oxman \\Instituto de F\'{\i}sica,
  Universidade Federal Fluminense, \\
  Campus da Praia Vermelha, Niter\'oi,\\ 24210-340, RJ, Brazil.}
\begin{document}
\maketitle
\begin{abstract}
  We study the consequences of including parity preserving matter 
  for the effective dual theory corresponding to
  compact $QED_3$; in particular we focus on the effect of that contribution
  on the confinement-deconfinement properties of the system.  To that end, we
  compare two recent proposals when massless fermions are included, 
  both based on an effective anomalous dual model, but having global and local
  $Z_2$ symmetries, respectively.
  
  We present a detailed analysis to show that while for large mass fermions
  the global $Z_2$ symmetry is preferred, in the massless fermion case the
  local $Z_2$ scenario turns out to be the proper one.
  
  We present a detailed discussion about how the inclusion of massless
  fermions in compact $QED_3$ leads to deconfinement, and discuss the
  stability of the deconfined phase by introducing a description based on an
  instanton dipole liquid picture.
  
\end{abstract}
\section{Introduction}
Three-dimensional non Abelian gauge theories have many properties that make
them very interesting from the point of view of the realization and
description of the confinement mechanism. Among those properties a remarkable
one is that, in the low-energy regime, there exist effective descriptions
which exhibit confinement in a very simple way, already in a semiclassical
approach. As explained by 't Hooft~\cite{thooft}, for the case of an $SU(N)$
gauge field theory, the natural degrees of freedom of the effective field
theory turn out to be {\em magnetic vortex\/} fields, with the confinement
mechanism being tantamount to the confinement of topological solitons for
those vortex fields. Using this approach, the phase structure of the $SU(N)$
Georgi-Glashow model in $2+1$ dimensions can be obtained by showing that the
$Z_N$ magnetic vortices condense, as in this situation the heavy electric
charges are confined. Indeed, ${\mathcal L}_V$, the effective Euclidean
Lagrangian for the dual $Z_N$ theory is given by
\begin{equation}\label{vconde}
  {\mathcal L}_V \;=\; \partial_\mu \bar{V}\partial_\mu V + \lambda
  (\bar{V}V - \mu^2)^M + \zeta (V^N +\bar{V}^N)\;,
\end{equation}
where $V$ is the vortex field operator whilst $\lambda$, $\mu$ and $\zeta$
denote constant parameters. $M$ is an integer bigger than $N$, to ensure that
the energy is bounded from below.

Vortex condensation corresponds to the spontaneous breaking of the magnetic
$Z_N$ symmetry in (\ref{vconde}). In this case, domain wall defects can be
formed, which for planar systems are, of course, one-dimensional objects. In
order to be observable, a pair of heavy quarks must be interpolated by this
kind of one-dimensional `string', what results in a configuration energy which
grows linearly with the pair separation distance~\cite{thooft}.

A concrete realization of this idea had been previously introduced for the
case of the $SU(2)$ Georgi-Glashow model~\cite{polyakov}.  In the London
limit, the relevant long-distance physics is dominated by the (unbroken)
$U(1)$ subgroup and, in that regime, the theory behaves as compact $QED_3$, in
the sense that it admits instantons. Far from their center, the instanton
gauge fields along the locally unbroken $U(1)$ direction look like Dirac
monopoles (where one of three coordinates is the Euclidean time), and their
action is UV finite since they have a finite size core. By using a singular
gauge transformation, the original (monopole-like) instanton configurations
can be also gauged along a fixed $U(1)$ subgroup (unbroken by the vacuum) to
be seen as Dirac monopoles in the whole space (with a corresponding Dirac
string). This defines an effective Abelian theory (see also~\cite{digiaco})
where the above mentioned finite size core may be interpreted as corresponding
to a minimum length (i.e., an UV cutoff), a feature one should have, since
compact $QED$ is usually defined on a spacetime lattice.

Thus, as explained in~\cite{polyakov}, a consequence of the compactness of the
underlying $SU(2)$ theory is the existence of (finite action) instanton
configurations, which for pure compact $QED_3$ are properly described by a
Coulomb gas.  The dual description is then given by the (Euclidean) Lagrangian
density,
\begin{equation}\label{lpolya}
   {\mathcal L}_{dual} \;=\; \frac{1}{2g_m^2} 
\partial_\mu \phi \partial_\mu \phi + \xi \mu^3 \cos \phi \;.
\end{equation}
A comparison with the $N=2$ model in (\ref{vconde}) shows that (\ref{lpolya})
represents a vortex model where the global $Z_2$ symmetry is spontaneously
broken (vortices condense), with the natural identification $V=V_0 \exp
(i\phi/2)$ (see also~\cite{kovner}). In terms of this dual model, Polyakov
obtained an area law for the Wilson loop in pure compact $QED_3$, a signal of
the confinement of electric charges.

Several authors have afterwards discussed the effects due to the
inclusion of parity preserving dynamical matter in the {\em fundamental\/}
representation, a nontrivial modification of the model,
since the $Z_N$ symmetry is derived under the assumption that
matter, when present, is in the adjoint representation.

In~\cite{kleinert,igor}, the effect of {\em massless\/} fermions was taken
into account by keeping just the quadratic part of their contribution to the
effective action, a procedure which is justified by invoking an IR fixed point
argument. As it is well known, when parity is not broken, the quadratic part
of the fermionic effective action contributes a nonlocal Maxwell term $S_{nl}$
to the effective action,
\begin{equation}\label{nl}
   S_{nl} \;=\; \frac{1}{2} \xi \int d^3 x  \; f_\mu
   \frac{1}{\sqrt{-\partial^2}} f_\mu \; ,
\end{equation}
where $f_\mu$ is the dual of the field strength tensor corresponding to the
gauge field $A_\mu$, namely, \mbox{$f_\mu \equiv \epsilon_{\mu\nu\lambda}
  \partial_\nu A_\lambda$}, and $\xi$ is a dimensionless constant which, for a
single fermionic flavour is $\xi=\frac{1}{16}$. In particular,
in~\cite{kleinert}, based on the instanton anti-instanton logarithmic
interaction implied by (\ref{nl}), arguments in favour of a deconfined phase
associated with instanton suppression were presented.

  In these references, compactness of the $U(1)$ model was simulated by
  introducing instanton gas configurations (whose action is decreed to be
  finite) into the anomalous compact $QED_3$ model based on (\ref{nl}).
  As a result, a dual description displaying a global $Z_2$ symmetry was
  obtained,
\begin{equation}\label{lapolya}
   S\;=\; \frac{1}{2g_m^2} \partial_\mu \phi (-\partial^2)^{1/2}
   \partial^\mu \phi + \xi \mu^3 \cos \phi \; ,
\end{equation}
which corresponds to an anomalous version of (\ref{lpolya}).

However, as we will see, the computation of the Wilson loop based on the
effective model (\ref{lapolya}) would give a destabilization of the area law,
instead of the perimeter law characterizing deconfinement. This can be traced
back to the global $Z_2$ symmetry of the anomalous dual action (\ref{lapolya})
which implies that the relevant source when computing the Wilson loop is
indeed concentrated on a surface.

On the other hand, in~\cite{cesar}, it had been argued that in the $SU(N)$
Georgi-Glashow model with dynamical matter, the global $Z_N$ becomes local,
implying that the Wilson operator changes the vacuum inside the associated
Wilson loop to a gauge equivalent state. Then, as discussed in \cite{cesar},
the overlap between the vacuum and the new state is expected to give a
boundary effect and, depending on the localization properties of the model, a
perimeter law would instead be obtained.

In~\cite{lucho}, compact $QED_3$ with dynamical matter was studied by using a
bosonized version of the fermion sector. In this way, the local $Z_2$ theory
invoked in~\cite{cesar} emerges naturally, with the bosonizing field
$\tilde{A}_\mu$ playing the role of the $Z_2$ gauge field. Moreover, in this
reference, the source which is relevant to compute the average of the Wilson
operator is shown to be concentrated on the loop and, for massless fermions,
arguments in favour of the perimeter law were presented.

The point here is that, even in the massive case, the local $Z_2$ is still
present. Thus, when the fermions become heavy, a destabilization
of the perimeter law is obtained, instead of the expected area
law.

The question suggests itself about how, by tuning the fermion mass, different
low energy phases arise.  When the dynamical fermion mass is large, fermions
are almost decoupled and the area law (confinement) corresponding to the
quenched theory should emerge. On the other hand, for light fermions a
perimeter law characterizing deconfinement is expected.  In this regard, more
recently, a U(1) 3D compact lattice gauge theory having nonlocal interactions
simulating the effect of the fermionic effective action has been studied
(see~\cite{j}).  Depending on whether a localized (representing massive
fermions) or nonlocalized interaction (representing massless fermions) is
considered, the fitting of the Wilson loop using the lattice data prefers the
area or the perimeter law, respectively.

Our main proposal here is to discuss how the fermion mass is reflected in the
corresponding low energy effective dual description of the continuum theory.
Once the proper effective description is established, we will determine the
corresponding Wilson loop behaviour, analyzing confinement vs deconfinement.

The structure of this article is as follows: in section~\ref{sec:ncwilson}, we
start by analyzing the main properties of the Wilson loop in low dimensional
noncompact gauge theories with and without dynamical fermions. This analysis
will turn out to be useful when discussing some compactification scenarios in
the presence of massless fermions. In section~\ref{sec:cwilson}, we will
compare the Wilson loop representation implied by the global and local $Z_2$
dual descriptions, respectively.

In particular, we will show that the local $Z_2$ symmetry amounts to the
complete erasure of instantons, that is, the theory becomes effectively
noncompact.

In section~\ref{zm}, we analyze the effect coming from quasi-zero modes
present in the zero charge instanton sector in compact $QED$ with massless
fermions, interpreting the result in terms of a binding interaction between
opposite instanton charges. In this manner, we discuss the physical conditions
that render the dual model with local $Z_2$ symmetry reliable, showing that
the inclusion of massless fermions in compact $QED_3$ drive the system into a
deconfined phase.

In section~\ref{idl}, we study possible deviations from the local $Z_2$
(noncompact) scenario introducing a novel description when instantons are not
completely suppressed: the instanton dipole liquid.  In this manner, we will
be able to discuss the stability of the deconfined phase.

Finally, we present our conclusions, comparing our approach with previous ones
discussed in the literature.

\section{Wilson loop in noncompact models}\label{sec:ncwilson}

A quite useful object for the study of confinement in gauge
theories is $\langle W({\mathcal C}) \rangle$, the vacuum average
of the Wilson loop operator $W({\mathcal C})$ corresponding to a
(smooth) closed curve ${\mathcal C}$, which (in $d$ spacetime
dimensions) is defined by
\begin{equation}
  W({\mathcal C})\;=\; e^{i\oint_{\mathcal C} dx_\mu A_\mu } \;.
\end{equation}
It can also be written in an equivalent form by introducing the dual of the
$F_{\mu\nu}$ field. For example, in $d=2$:
\begin{equation}
  W({\mathcal C})\;=\; e^{i \int d^2x \,\chi_\Sigma \, f }\;  ,
\end{equation}
where $f(x) = \frac{1}{2} \epsilon_{\mu\nu}F_{\mu\nu}(x)$ and $\chi_\Sigma(x)$
is the characteristic function of the plane surface $\Sigma$, enclosed by
${\mathcal C}$. In $d=3$ we have instead:
\begin{equation}\label{ws}
 W({\mathcal C}) \;=\; e^{i\int d^dx\, s_\mu\, f_\mu }\;,\;\;
  s_\mu\;\equiv\;g\,\delta_\Sigma\, n_\mu\; ,
\end{equation}
where $\delta_\Sigma$ is a Dirac distribution with support on a spacetime
surface \mbox{$\Sigma = \Sigma({\mathcal C})$} whose boundary is the curve
${\mathcal C}$, and $n_\mu$ is the field of unit vectors normal to $\Sigma$,
defined by the surface element $dS_\mu=n_\mu\, dS$.

The vacuum average $\langle \ldots \rangle$ is evaluated differently depending
on whether the theory includes or not matter fields. In the latter
(`quenched') case
\begin{equation}\label{eq:defav1}
  \langle \ldots \rangle \;=\; \frac{\int {\mathcal D}A \;\ldots
  \; e^{-S_G [A]}}{\int {\mathcal D}A \; e^{- S_G [A]}}\; ,
\end{equation}
where $S_G[A]$ is the gauge field action (we omit writing gauge fixing objects
explicitly, since we assume them to be part of the measure ${\mathcal D}A$).
When charged matter is included, we have instead:
\begin{equation}\label{eq:defav2}
  \langle \ldots \rangle \;=\; \frac{\int {\mathcal D}A \;\ldots\;
  e^{-S_G [A] - \Gamma_F[A]}}{\int {\mathcal D}A \; e^{- S_G [A]- \Gamma_F[A]}}\; ,
\end{equation}
where $\Gamma_F[A]$ is the contribution of the matter (fermion) fields to the
gauge field effective action:
\begin{equation}\label{eq:defgeff}
    e^{- \Gamma_F[A]} \;=\; \int \,{\mathcal D}{\bar\psi} \,{\mathcal
    D}\psi \; e^{- S_F({\bar\psi},\psi,A)}\; ,
\end{equation}
with
\begin{equation}\label{eq:defsf}
    S_F({\bar\psi},\psi,A) \;=\; \int d^dx \,{\bar\psi} {\mathcal D} \psi
\end{equation}
and ${\mathcal D}= \not \!\partial + i \not \!\! A + m$\;.

In the two following subsections we evaluate the Wilson loop in noncompact
$QED_2$ and $QED_3$ discussing the effect of introducing dynamical fermions.
Let us begin with $QED_2$, a warming-up exercise, where everything may be
exactly calculated (at least when the fermions are massless).
\subsection{Noncompact $QED_2$}\label{ssec:qed2}
Let us first consider $\langle W({\mathcal C}) \rangle$ for the
quenched {\em noncompact\/} theory:
\begin{equation}\label{eq:defwav}
\langle W({\mathcal C}) \rangle \;=\;\frac{\int {\mathcal D}A
  \; e^{-S_G [A] + i \int s A}}{\int {\mathcal D}A \; e^{- S_G [A]}}\; ,
\end{equation}
where $S_G \equiv \frac{1}{2 g^2} \int d^2x f^2$, and $g$ is the
coupling constant. We then introduce an auxiliary pseudoscalar
field $\lambda (x)$ to write the equivalent expression:
\begin{equation}\label{eq:auxf1}
\langle W({\mathcal C}) \rangle \;=\;\frac{\int {\mathcal
D}\lambda \, {\mathcal D}A \; e^{- \int \frac{g^2}{2}\lambda^2 + i
\int  f (\lambda + \chi_\Sigma)}}{\int {\mathcal D}\lambda \,
{\mathcal D}A \; e^{- \int \frac{g^2}{2} \lambda^2 + i \int f
\lambda }} \;.
\end{equation}
Integrating $A_\mu$ yields the constraint \mbox{$\lambda + \chi_\Sigma =
  \kappa$}, where $\kappa$ is a constant. Its value is fixed to zero by using
the boundary condition \mbox{$\lambda (x) \to 0$} for \mbox{$x \to \infty$}.
Using this constraint into (\ref{eq:auxf1}) yields
\begin{equation}
    \langle W({\mathcal C}) \rangle \;=\; e^{- \frac{g^2}{2} \int d^2x \, 
(\chi_\Sigma)^2}
    \;=\;e^{- \frac{g^2}{2} \int d^2x \, \chi_\Sigma}\;.
\end{equation}
Finally, if ${\mathcal A}_\Sigma$ denotes the area of $\Sigma$ (the area
enclosed by ${\mathcal C}$) we have just shown that:
\begin{equation}
    \langle W({\mathcal C}) \rangle \;=\; 
e^{- \frac{g^2}{2} \, {\mathcal A}_\Sigma }\; ,
\end{equation}
as it should be, since we know that the model is confining.

When a Dirac matter field is introduced, we have to add the effective action
contribution, which for massless fields is given by:
\begin{equation}\label{eq:unq2}
    \Gamma_F (A) \;=\; (2\pi)^{-1}  \; \int d^2x \; f
    \, \frac{1}{(-\partial^2)} \, f\;.
\end{equation}
An entirely analogous calculation to the previous one yields now the result:
\begin{equation}
    \langle W({\mathcal C}) \rangle \;=\; e^{- \frac{g^2}{2} \int \,
    \chi_\Sigma \,{\mathcal O}^{-1} \,\chi_\Sigma}\; ,
\end{equation}
where
\begin{equation}
{\mathcal O} \;=\; 1 \,-\, \frac{g^2}{\pi} \, \partial^{-2}\;.
\end{equation}

For the particular case of a circular curve of radius $R$, the previous
expression may be exactly calculated. Indeed, by introducing the Fourier
representation for the kernel of ${\mathcal O}^{-1}$,
\begin{equation}\label{eq:ofourier}
    {\mathcal O}^{-1}(x,y) \;=\; \int \frac{d^2k}{(2\pi)^2} \,
    e^{i k \cdot (x-y)} \; \frac{1}{1 + \frac{g^2}{\pi k^2}}\;,
\end{equation}
we see that
\begin{equation}
\int \,\chi_\Sigma {\mathcal O}^{-1} \chi_\Sigma \;=\; \int
\frac{d^2k}{(2\pi)^2} \; \frac{|{\tilde \chi}(k)|^2}{1 +
\frac{g^2}{\pi k^2}}\; ,
\end{equation}
where
\begin{equation}
    {\tilde \chi}(k) \;=\; \int d^2x \, \chi_\Sigma (x) \, e^{i k
    \cdot x} \;=\; \frac{2 \pi R}{k} \, J_1 (k R) \;,
\end{equation}
with $J_1$ denoting a Bessel function of the first kind. Then
\begin{equation}
    \ln \Big[ \langle W({\mathcal C}) \rangle \Big]\;=\; -
    \frac{g^2}{2} \; R^2 \, \int \, d^2k \; \frac{\big(J_1(k)\big)^2}{k^2 +
\frac{g^2}{\pi} R^2} \;,
\end{equation}
which can be exactly calculated:
\begin{equation}
    \ln \Big[ \langle W({\mathcal C}) \rangle \Big]\;=\; - \, \pi
    \, g^2 \; R^2  \; I_1(\frac{g}{\sqrt{\pi}} R) \;
    K_1(\frac{g}{\sqrt{\pi}}R) \;.
\end{equation}
Here, $I_1$ and $K_1$ are modified Bessel functions. When $R$ is very large
($\frac{g}{\sqrt{\pi}}R >> 1$), we may use the asymptotic
form~\footnote{Next-to-leading terms are exponentially small.}:
\begin{equation}
 I_1(x) \;K_1(x) \;\sim \;\frac{1}{2 \, x} \;\;\; ( x
 >> 1)\;,
 \end{equation}
 to see that
\begin{equation}
    \ln \Big[ \langle W({\mathcal C}) \rangle \Big]\;=\; - \,
    \frac{\pi^{\frac{1}{2}}}{4}\, g \times  2 \pi R \;,
\end{equation}
i.e., a perimeter law, a clear signal of the fact that the
massless fermions drive the system into a deconfined phase. Of
course, the area law is recovered if the $g \to 0$ asymptotic form
is taken for the Bessel functions.

\subsection{Noncompact $QED_3$}\label{ssec:qed3}
As in the $QED_2$ case, we first consider the quenched {\em noncompact\/}
theory. Of course, here we have an expression for the Wilson-loop average
which is formally equal to the one introduced in (\ref{eq:defwav}), except for
the necessary changes due to the different number of spacetime dimensions.
Namely, the gauge-field action $S_G$ is now:
\begin{equation}
S_G \;\equiv \;\frac{1}{2 g^2} \int d^3x f_\mu f_\mu \;\;, \;\;\;\;
f_\mu \;=\; \epsilon_{\mu\nu\rho} \partial_\nu A_\rho \;,
\end{equation}
while the term corresponding to the `coupling' of the loop with $A_\mu$,
\mbox{$\int d^3x s_\mu f_\mu$}, deserves a closer look. To write it in a more
explicit fashion, we first note that the $\delta_\Sigma$ distribution may be
given a more concrete form by introducing a parametrization $X_\mu$ of the
surface $\Sigma({\mathcal C})$, bounded by the curve ${\mathcal C}$:
\begin{equation}\label{eq:sigmapar}
X_\mu : \; [0,1] \times [0,1] \;\to\; {\mathbb R}^3 \; ,
\end{equation}
where $X_\mu (\sigma^1, \sigma^2)$ is a parametric representation of $\Sigma$.
The two worldsheet coordinates $\sigma^a$, $a=1,2$ are to some extent
arbitrary, as well as their domain of definition. Nevertheless,
$\delta_\Sigma$, must of course be reparametrization invariant.

Then we see that $\delta_\Sigma$ is:
\begin{equation}\label{eq:deltapar}
\delta_\Sigma (x) \;=\; \int_0^1 d\sigma^1 \int_0^1 d\sigma^2 \;
\sqrt{g(\sigma)} \; \delta^{(3)}\big[ x \,- \,X(\sigma) \big] \; ,
\end{equation} 
where $g(\sigma)$ denotes the determinant of $g_{ab}$, the induced metric on
the surface:
\begin{equation}
g(\sigma) \;\equiv\; \det \big[ g_{ab} (\sigma) \big] \;\;,\;\;\;
g_{ab}(\sigma) \;=\; \frac{\partial X_\mu}{\partial \sigma^a} 
\frac{\partial X_\mu}{\partial \sigma^b} \;.
\end{equation}
The $\sqrt{g(\sigma)}$ factor is of course required in order to have an
invariant volume on the worldsheet, and as a consequence a
reparametrization-invariant definition for $\delta_\Sigma$.
To complete the construction of $s_\mu$, we have to give an expression for
${\hat n}_\mu (\sigma)$, the normal to $\Sigma$ at the point parametrized by
$\sigma$. The surface element $dS_\mu (\sigma)$ is:
\begin{equation}
dS_\mu (\sigma) \;=\; \epsilon_{\mu\nu\rho} \frac{\partial X_\nu}{\partial
  \sigma_1} \frac{\partial X_\rho}{\partial \sigma_2}\;, 
\end{equation}
which points in the normal direction, but is not normalized. Hence:
\begin{equation}
{\hat n}_\mu (\sigma) \;\equiv\; \frac{dS_\mu (\sigma)}{|d S (\sigma)|}   \;, 
\end{equation}
and an elementary calculation shows that:
\begin{equation}
|d S | \;\equiv \; \sqrt{dS_\mu dS_\mu} \;=\; \sqrt{\det (g_{ab})} \;=\;
  \sqrt{g(\sigma)} \;. 
\end{equation}
Thus:
\begin{equation}
s_\mu (x) \;=\;  \int_0^1 d\sigma^1 \int_0^1 d\sigma^2 \;
\sqrt{g(\sigma)} \; \delta^{(3)}\big[ x \,- \,X(\sigma) \big] \; 
{\hat n}_\mu(\sigma) \; ,
\end{equation}
and using the explicit form of ${\hat n}_\mu$, we obtain
\begin{equation}
s_\mu (x) \;=\;  \int_0^1 d\sigma^1 \int_0^1 d\sigma^2 \;
\delta^{(3)}\big[ x \,- \,X(\sigma) \big] \; 
 \epsilon_{\mu\nu\rho} \frac{\partial X_\nu}{\partial
  \sigma_1} \frac{\partial X_\rho}{\partial \sigma_2} \;.
\end{equation}
Now an auxiliary {\em pseudovector\/} field $\lambda_\mu (x)$ is introduced in
order to write an equivalent expression where $A_\mu$ only appears linearly in
the exponent:
\begin{equation}\label{eq:auxf1}
\langle W({\mathcal C}) \rangle \;=\;\int {\mathcal
D}\lambda \, {\mathcal D}A \; e^{- \frac{g^2}{2} \int \lambda_\mu^2 + i
\int  f_\mu  (\lambda_\mu + s_\mu)} 
\end{equation}
(we have absorbed the normalization factor into the measure).  Again, the
integral over $A_\mu$ yields a linear constraint: \mbox{$\lambda_\mu + s_\mu =
  \partial_\mu \phi$}, where $\phi$ is an (undetermined) pseudoscalar field.
Then $\lambda$ can be completely integrated out by using that constraint, but
there still remains an integration over $\phi$:
\begin{equation}
    \langle W({\mathcal C}) \rangle \;=\; \int {\mathcal D}\phi \; 
\exp\Big[ - \frac{g^2}{2} \int d^3x \,\big( s_\mu - \partial_\mu \phi\big)^2
\Big] \;. 
\end{equation}
It should be clear that the effect of the integration over $\phi$ is to erase
any dependence of the average on the longitudinal part of $s_\mu$. Indeed,
\begin{eqnarray}\label{eq:lnwqnc}
   \ln \Big[ \langle W({\mathcal C}) \rangle \Big] &=& 
- \frac{g^2}{2} \int d^3x \,s_\mu(x) s_\mu(x) \nonumber\\
&+& \frac{1}{2} \, g^2 \, \int d^3x \int d^3y\; \partial \cdot s(x) \,
G(x-y) \, \partial \cdot s(y) \Big] \;, 
\end{eqnarray}
where 
\begin{equation}
G(x-y) \;=\; \int \frac{d^3k}{(2\pi)^3} \, \frac{e^{i k \cdot (x-y)}}{k^2} 
\;=\; \frac{1}{4 \pi \,|x-y|} \;.
\end{equation}
The divergence of $s_\mu$ is obviously concentrated on the surface $\Sigma$,
and there is a non-trivial (partial) cancellation between the
two terms on the rhs of (\ref{eq:lnwqnc}). To see that more clearly, we
consider the simple (and relevant) case of a planar curve, contained in the
plane $x_3=0$:
\begin{equation}
s_\mu (x) \;=\; \delta_{\mu 3} \, \delta(x_3)\, \chi_\Sigma ({\mathbf x})\; ,
\end{equation}
where ${\mathbf x}$ denotes the two coordinates on the $x_3$ plane.  Since
$\partial \cdot s (x) = \delta'(x_3) \chi_\Sigma ({\mathbf x})$, an elementary
calculation shows that the contribution of the $s_\mu^2$ term (which contains
a $\delta(0)$) is cancelled by a contribution from the second term, but there
is a non-vanishing remainder:
\begin{equation}
   \ln \Big[ \langle W({\mathcal C}) \rangle \Big]\;=\; 
\frac{1}{2} \, g^2 \, \int d^2x \int d^2y\; \chi_\Sigma ({\mathbf x}) 
\big[\frac{\partial^2}{\partial {\mathbf x}^2} G({\mathbf x}- {\mathbf y},
x_3)\big]_{x_3 = 0}    \chi_\Sigma ({\mathbf y}) \;.
\end{equation}
Or, in terms of $\tilde{\chi}_\Sigma$, the (two-dimensional) Fourier transform
of $\chi_\Sigma$,
\begin{equation}
   \ln \Big[ \langle W({\mathcal C}) \rangle \Big]\;=\; - \, 
\frac{1}{2} \, g^2 \, \int \frac{d^2k}{(2\pi)^2} \,\frac{|{\mathbf k}|}{2} \,
| {\tilde \chi}_\Sigma ({\mathbf k}) |^2 \;.
\end{equation}
We note that the cancellation between two contributions in (\ref{eq:lnwqnc})
gets rid of a divergent term proportional to the area of $\Sigma$.

To derive concrete results for $\langle W\rangle$, it is convenient to assume
a particular form for the curve ${\mathcal C}$ (and its associated surface
$\Sigma$). When ${\mathcal C}$ is a circle of radius $R$, using ${\tilde
  \chi}({\mathbf k})=\frac{2 \pi R}{k} \, J_1 (k R)$, we see that:
\begin{equation}
   \ln \Big[ \langle W({\mathcal C}) \rangle \Big]\;=\; -\,
\frac{\pi}{2} \, g^2 \, R \;  \int_0^{\Lambda R} du \, [ J_1 (u) ]^2 \;,
\end{equation}
where we have introduced an Euclidean UV cutoff, $\Lambda$.  The integral over
$u$ can be exactly calculated; this yields
\begin{equation}
   \ln \Big[ \langle W({\mathcal C}) \rangle \Big]\;=\; - \,
\frac{\pi}{24} \, g^2 \, R \; (\Lambda R)^3 \, _2F_3 \big(\frac{3}{2}
,\frac{3}{2};2,\frac{5}{2},3;-(\Lambda R)^2 \big) \;,
\end{equation}
where $_2F_3$ is a generalized hypergeometric function.  The last expression
does not allow for a clean analysis of the dependence of the loop average with
the distance, for long distances.  This problem may be solved by using a
milder form of regularization: let us consider the regularized function
\begin{equation}
 \ln \Big[ \langle W({\mathcal C}) \rangle \Big]_{reg}  \;\equiv\;
 - \, \frac{1}{2} \, g^2 \, \mu^{1 - \alpha} \; \int \frac{d^2k}{(2\pi)^2} \,
  \frac{|{\mathbf k}|^\alpha}{2} \, 
|{\tilde \chi}_\Sigma ({\mathbf k}) |^2 \;,
\end{equation}
with the obvious regularization parameter $\alpha$. Note that, in order to
keep the result dimensionless, we have had to introduce a constant $\mu$, with
the dimensions of a mass. The regularization is removed by letting $\alpha = 1
- \varepsilon$, with $\varepsilon \to 0$, and UV divergences emerge as poles
in the complex variable $\varepsilon$.

Now the integral over ${\mathbf k}$ may be exactly performed for the
analytically continued function, with the result:
\begin{equation}
 \ln \Big[ \langle W({\mathcal C}) \rangle \Big]_{reg}  \;=\;
 \frac{\pi}{2} \, g^2 \, R \; (\mu R)^{1-\alpha} \, 
\frac{\Gamma(\frac{1}{2} - \frac{\alpha}{2}) \; \Gamma(1 +
 \frac{\alpha}{2})}{\sqrt{\pi} \;\alpha \;
\Gamma(-\frac{\alpha}{2}) \; \Gamma(2 - \frac{\alpha}{2})} \;.
\end{equation}
Now we set $\alpha = 1 - \varepsilon$ and collect the relevant terms when
$\varepsilon \to 0$, obtaining:
\begin{equation}
 \ln \Big[ \langle W({\mathcal C}) \rangle \Big]_{reg}  
\;\sim \;- \frac{1}{2} \, g^2 \; \big[ \frac{1}{\varepsilon} R  \,+\, 
 R \; \ln(\mu R) \big] \;\;\; (\varepsilon \sim 0) \;.
\end{equation}
The $\varepsilon^{-1}$ term is of course divergent when the regularization is
removed. A subtracted version of the average will then have the general form:
\begin{equation}
 \ln \Big[ \langle W({\mathcal C}) \rangle \Big]_{sub}  
\;\sim \;- \frac{1}{2} \, g^2 \; \big[ \kappa  R  \,+\, 
 R \; \ln(\mu R) \big] \;,
\end{equation}
where the (finite) constant $\kappa$ is to be fixed by a renormalization
condition. Of course, $\kappa$ may be absorbed into a different choice of the
dimensionful constant $\mu'$, and so we may use the expression
\begin{equation}
 \ln \Big[ \langle W({\mathcal C}) \rangle \Big]_{sub}  
\;\sim \;- \frac{1}{2} \, g^2 \; \big[ R  \; \ln(\mu' R) \big] \;,
\end{equation}
with a different $\mu'$, to be fixed, for example, by setting the value of the
average to zero at some small radius $a$, what yields the renormalized
average:
\begin{equation}\label{eq:wcirc3}
 \ln \Big[ \langle W({\mathcal C}) \rangle \Big]_{ren}  
\;\sim \;- \frac{1}{2} \, g^2 \; \big[ R  \; \ln(\frac{R}{a}) \big] \;,
\end{equation}
which is the result for the noncompact quenched case. Note that this result is
consistent with the one obtained for a rectangular loop and using dimensional
regularization~\cite{zinn}, where the static energy between point charges in
$d > 2$ is found to be:
\begin{equation}\label{eq:reszinn}
E(R) - E(a) \;=\; \frac{g^2}{4 \pi^{\frac{d-1}{2}}} \,\Gamma(\frac{d-3}{2}) \;
\big(a^{3-d} - R^{3-d}\big) \;,
\end{equation}

Taking the $d\to 3$ limit yields:
\begin{equation}\label{eq:reszinn1}
\Big[E(R) - E(a)\Big]_{d\to 3} \;=\; \frac{g^2}{2 \pi} \; \ln(\frac{R}{a}) \;,
\end{equation}
to be compared with: 
\begin{equation}\label{eq:circular}
E(R) - E(a) \;=\; \frac{g^2}{2} \; \ln(\frac{R}{a}) \;,
\end{equation}
for the circular loop, as follows from (\ref{eq:wcirc3}).

Things are quite different when one includes fermionic matter into the game,
since this adds the extra contribution already mentioned in equation
(\ref{nl}) of the Introduction: $\Gamma_F (A) \;=\; S_{nl}(A)$.

Keeping only this contribution has the obvious advantage of having a quadratic
gauge-field action. Thus, with just a few calculations we obtain now the
Wilson loop average in terms of the Fourier transform of $\chi_\Sigma$ for
this case also:
\begin{equation}
 \ln \Big[ \langle W({\mathcal C}) \rangle \Big]_{reg}  \;\equiv\;
 - \, \frac{1}{2} \, g^2 \, \mu^{1 - \alpha} \; \int \frac{d^3k}{(2\pi)^3} \,
  \frac{|{\mathbf k}|^{1 + \alpha}}{k^2} \; 
\frac{1}{1 + \xi \frac{g^2}{k}} \, 
|{\tilde \chi}_\Sigma ({\mathbf k}) |^2 \;,
\label{wncf}
\end{equation}
where we adopted the same analytic regularization as for the quenched case.

Rather than evaluating directly this expression, we shall consider the results
it yields in different limits. Of course, the $\xi \to 0$ limit is regular, in
the sense that we recover the behaviour of the quenched case:
\begin{equation}\label{eq:wcirc3f}
 \ln \Big[ \langle W({\mathcal C}) \rangle \Big]_{ren}  
\;\to \;- \frac{1}{2} \, g^2 \; \big[ R  \; \ln(\frac{R}{a}) \big] \;\;
\;\;\; \xi \to 0 \;.
\end{equation}
An interesting departure from this result arises when one considers the
would-be `large-$\xi$' limit, to be understood in the sense that only the
IR-dominant contribution from the quadratic operator is kept. This is, indeed,
justified for the calculation of the Wilson loop for a rectangular loop if one
is only interested in the large-$R$ case (in this case, $g^2 R >>1$. Then one
has:
\begin{equation}
 \ln \Big[ \langle W({\mathcal C}) \rangle \Big]_{reg}  \;\sim\;
 - \, \frac{1}{2 \xi} \, \mu^{1 - \alpha} \; 
\int \frac{d^3k}{(2\pi)^3} \,
\frac{|{\mathbf k}|^{\alpha+1}}{k} \, |{\tilde \chi}_\Sigma ({\mathbf k}) |^2 \;,
\end{equation}
which is of course independent of $g$.

This integral can be directly evaluated, or its value can be deduced from
known results corresponding to the $d=4$ case, since the Feynman gauge
propagator in this limit is identical to the one for $d=4$, except from a
constant factor. Indeed, one has:
\begin{equation}
\langle x | \frac{1}{\sqrt{-\partial^2}} | y \rangle \;=\; 
\frac{8}{\pi^2\, | x - y|^2} \;\;\; (d=3)\;, 
\end{equation} 
while
\begin{equation}
\langle x | \frac{1}{(-\partial^2)} | y \rangle \;=\; 
\frac{2}{\pi^2\, | x - y|^2} \;\;\; (d=4)\;, 
\end{equation}
and one is only interested in their values on a plane curve, so having a
different number of components is irrelevant. For a rectangular loop, we
obtain the result:
\begin{equation}
\ln \Big[\langle W({\mathcal C})\rangle \Big]  \;\sim\; 
\frac{g^2}{\pi \xi} \; \big(\frac{R}{a} \,-\, 1 \big) \;\;\; R g^2 >> 1 \;,
\label{behf}
\end{equation}
which shows a dramatic difference with the quenched case: the confining energy
becomes a {\em constant\/} when massless fermions are included. 

Note that, as expected, (\ref{eq:wcirc3}) is associated with the
logarithmic interaction between electric charges, typical of planar noncompact
$QED_3$, while the behaviour in (\ref{behf}) corresponds to a $1/R$
interaction between well separated electric charges. This comes about as, at
large distances, (\ref{wncf}) is dominated by the action $S_{nl}(A)$ given
in (\ref{nl}), and for planar systems, this kind of nonlocal action is
known to imply a $1/R$ (Coulomb) static potential between electric charges
(see~\cite{BG}).

\section{Compactification scenarios with fundamental matter}\label{sec:cwilson}

\subsection{Instanton gas and $QED_3$: Global $Z_2$ symmetry}
\label{ssec:mg}

Let us now consider the compact theory with a general nonlocal action.
We shall consider the functional average:
\begin{equation}
\langle W({\mathcal C}) \rangle \;=\;\int {\mathcal D}j \, {\mathcal D}A \; 
e^{- \frac{1}{2 g^2} \, \int  (f + j) {\mathcal K} (f + j) + 
i \int s  (f + j)} \; ,
\end{equation}
where, to account for the neutral instanton background, we have made the
substitution:  
\begin{equation}\label{reple}
f_\mu (x)  \rightarrow f_\mu (x) + j_\mu (x)\;,\;\;\;
\end{equation}
with
\begin{equation}
j_\mu (x)\;=\; g_m\int_\gamma dy_\mu\, \delta^{(3)} (x-y),\;\;\; g\, g_m =2\pi,
\end{equation}
where $\gamma$ is a set of Dirac strings associated with instanton
anti-instanton (monopole-like) singularities located at the string endpoints.
We have used a compact notation in the exponent, so that, for example,
\begin{equation}
\int f {\mathcal K} f \;\equiv\; \int d^3x d^3y f_\mu(x) {\mathcal K}(x,y)
  f_\mu(y)\;,
\end{equation}
etcetera. Here, ${\mathcal K}(x,y)$ is the kernel of an operator which, in
the coordinate representation, is in general nonlocal. 

After introducing the auxiliary field $\lambda$ and integrating over the
$A_\mu$ field, we obtain
\begin{equation}
\langle W({\mathcal C})\rangle \;=\; \int {\mathcal D}j\, {\mathcal D}\phi\,
e^{-\int \,  \frac{1}{2} (\partial\phi + s ){\mathcal K}^{-1}
(\partial\phi + s ) + i\, \int j_\mu \partial_\mu \phi}\;.
\label{ii}
\end{equation}
Now, if an integration over a dilute monopole plasma is supposed,
\begin{eqnarray}
\int {\mathcal D}j\, e^{i\int \, \phi \partial_\mu j_\mu}&=&\sum_N
\sum_{q_a=\pm g_m}
\zeta^N/N! \int dz_1 \dots dz_N\, \prod_a e^{iq_a \phi (z_a)},\nonumber \\
&=&\sum_N \zeta^N/N! \left\{ \int dx\, 
(e^{ig_m\phi(x)}+e^{-ig_m\phi(x)})\right\}^N,
\end{eqnarray}
the following representation is obtained,
\begin{equation}
\langle W({\mathcal C})\rangle \;=\; \int {\mathcal D}\phi\, 
e^{-\int d^3x\,\big[ \frac{1}{2g_m^2}
(\partial\phi + g_m s ){\mathcal K}^{-1}(\partial\phi + g_m s )
+ 2 g_m^2\zeta \cos \phi \big] }\;,
\label{dp}
\end{equation}
where the field redefinition $g_m\, \phi \rightarrow \phi$ and Dirac's
quantization condition $g\, g_m=2\pi$ has been considered.

In the local case, ${\mathcal K}(x,y)\propto \delta(x-y)$ and (\ref{dp}) is
just Polyakov's dual representation for the Wilson loop for pure compact
$QED_3$.  According to Polyakov's analysis, the saddle point is given by a
domain wall containing a $2\pi$ discontinuity at $\Sigma$, coming from the
source term.  Therefore, when computing $(\partial\phi +g_m s)^ 2$ in
(\ref{dp}), the associated delta singularity in $\partial \phi$ cancels
against the $g_m s=2\pi\,\delta_{\Sigma}\,n$ term. Because of the localization
scale, away from $\Sigma$, the field is exponentially suppressed. Then, for a
large surface, the domain wall leads to an almost uniform action density
localized on $\Sigma$ implying the area law for the Wilson loop, that is,
linear confinement. Note that this analysis would also apply to the case where
large mass fermionic matter is included, as in this case the fermionic
effective action would contribute with just a local Maxwell term.

On the other hand, when massless fermions are present, including the result
from the one-loop contribution given by $S_{nl}$ (cf. (\ref{nl})) in the
quadratic approximation, we see that
\begin{eqnarray}
{\mathcal K}(x,y) &=& 
\langle x |\big[ 1 + \frac{ \xi g^2}{\sqrt{-\partial^2}} \big] |y\rangle
\nonumber\\ 
&=& \int \frac{d^3k}{(2\pi)^3} \,e^{ i k \cdot (x-y)}\, 
\big(1 + \frac{ \xi g^2}{k} \big) \;.
\label{Kmf}
\end{eqnarray}

At large distances, the second term in (\ref{Kmf}) dominates over the initial
pure local Maxwell term. In this case, when disregarding the source terms in
(\ref{dp}), the anomalous sine-Gordon model considered in~\cite{kleinert,igor}
(cf. (\ref{lapolya})) is obtained. As in this model the relevant source to
compute the Wilson loop continues to be concentrated on the surface $\Sigma$
(cf. (\ref{dp})), or in other words, the $Z_2$ symmetry is still global, a
perimeter law is not established in this framework, as required by a
deconfining phase, but rather a destabilization of the area law due to the
nonlocal kinetic operator.

\subsection{Local $Z_N$ and the Wilson loop}

In~\cite{cesar}, by using the vortex condensation approach to the $(2+1)$D
Georgi-Glashow model, the gauging of the $Z_N$ symmetry in (\ref{vconde}) has
been proposed, when matter in the fundamental representation is included.  In
this manner, a simple argument to discuss the Wilson loop behavior was
presented. We review it here: In the absence of dynamical quarks, the
(spontaneously broken) $Z_N$ symmetry is global.  The VEV of $W({\mathcal C})$
is the overlap of the vacuum state $|0>$ with the state $|S>$, which is
obtained from the vacuum by acting on it with the operator $W({\mathcal C})$.
When acting on the vacuum, $W({\mathcal C})$ performs a $Z_N$ transformation
at all points within the area $\Sigma$ bounded by ${\mathcal C}$.  If the
vacuum wavefunction depends on the configuration of non $Z_N$-invariant
degrees of freedom (the state in question is not $Z_N$ invariant) the action
of $W({\mathcal C})$ affects the state everywhere inside the loop.  On the
other hand, in the local theory with finite correlation length the overlap
between the two states approximately factorizes into the product of the
overlaps taken over the region of space of linear dimension of order of the
correlation length $l$
\begin{equation}
<0|\Sigma>=\Pi_x<0_x|\Sigma_x>\; ,
\label{fact}
\end{equation}
where the label $x$ is the coordinate of the point in the center of a given
small region of space. For $x$ outside $\Sigma$, the two states $|0_x>$ and
$|S_x>$ are identical and therefore their overlap is unity. However for $x$
inside $\Sigma$ the states are different and the overlap is therefore some
number $e^{-\gamma} < 1$. The number of such regions inside the area is
obviously of order $S/l^2$ and thus
\begin{equation}
<W({\mathcal C})>=\exp\{-\gamma{{\mathcal A}_\Sigma\over l^2}\}\;.
\end{equation}
The VEV of $W({\mathcal C})$ then falls off as an area.

On the other hand, when $Z_N$ is local, the action of a Wilson loop of finite
size inside the contour corresponds to a gauge transformation.  Thus in
physical terms locally inside the contour the new state is {\it the same\/} as
the old one and so the overlap between the two locally must be equal to one.
The only nontrivial contributions to the overlap may come from the region
close to the contour, thus (depending on the localization properties of the
theory) a perimeter law is obtained in this case.

\subsection{Bosonization and compactification of $QED_3$: Local $Z_2$ symmetry}
\label{boscom}

In~\cite{lucho}, before compactifying $QED_3$ with dynamical fermions, the
bosonized version of the matter sector was considered (see~\cite{bos1,bos2}),
\begin{equation}
K_F[\psi]+ig\int d^3x\, A^\mu J_\mu \leftrightarrow
K_B[\tilde{A}]+ig\int d^3x\, A^\mu \varepsilon _{\mu \nu \rho }\partial ^\nu \tilde{A}^\rho\; ,
\label{bos}
\end{equation}
where $K_F$ is the fermion action, and the $U(1)$ fermion matter current is
bosonized using a vector field $\tilde{A}_\mu$ whose gauge invariant action
$K_B$ is given by the transverse Fourier transform of the fermionic effective
action.  The relation in (\ref{bos}) is exact, and refers to the equivalence
of the partition functions associated with the fermionic field $\psi$ and the
bosonizing gauge field $\tilde{A}$, respectively. As discussed in
\cite{lucho}, the compactification process in this framework implies that the
$Z_2$ symmetry present in (\ref{lpolya}) becomes local. In this case, the
Wilson loop representation is (see~\cite{lucho}),
\begin{equation}
W(s)=\frac{1}{{\mathcal N}}\int {\mathcal D}\tilde{A}\, 
{\mathcal D}\phi\, e^{-\int d^3x\, \left(\frac{1}{2g_m^2}
(\partial\phi + g_m s - 2\pi \tilde{A} )^2
- \xi \mu^3 \cos \phi \right) - K_B[\tilde{A}]}\;.
\label{wdual}
\end{equation}
As the dual action $K_B[\tilde{A}]$ is gauge invariant for any associated
transformation $\tilde{A}\rightarrow \tilde{A}-(1/2\pi)\partial\phi$, the
$\phi$ field in the correlator (\ref{wdual}) decouples. In addition, parity
symmetry implies that $K_B[\tilde{A}]$ does not contain a Chern-Simons term;
it only depends on the dual tensor $\epsilon \partial \tilde{A}$. Then,
transforming $\tilde{A}\to \tilde{A}+ (g_m/2\pi) s$, the exponent in
(\ref{wdual}) contains the term $\int d^3x\, \frac{g^2}{2} \tilde{A}^2$, plus
the bosonized action which now depends on $\epsilon \partial
\tilde{A}+(g_m/2\pi)\epsilon \partial s$.

This led to a qualitatively different situation when compared with eq.
(\ref{dp}), where the relevant source $s_\mu$ for the Wilson loop is
concentrated on a surface.  The local $Z_2$ symmetry changes the relevant
source to be $\epsilon \partial s|^\mu=\epsilon^{\mu \nu \rho}n_\rho
\partial_{\nu}\delta_\Sigma$, which is tangent to the boundary of $\Sigma$ and
is concentrated there.

This change in the dimensionality of the source indicates that, depending on
the localization properties in (\ref{wdual}), a perimeter law could be
observed in the phase where dynamical quarks are coupled.

In order to evaluate the Wilson loop in the local $Z_2$ theory, gauging out
the field $\phi$, we obtain,
\begin{eqnarray}
W(s)&=&\frac{1}{{\mathcal N}}\int {\mathcal D}\tilde{A}\, e^{ -\int d^3x\,
\frac{1}{2}(s - g \tilde{A} )^2 - K_B[\tilde{A}]}\nonumber \\
&=& \frac{1}{{\mathcal N}}\int {\mathcal D}A\, {\mathcal D}\tilde{A}\, e^{ -\int d^3x\,
\frac{1}{2}f^2+i\int d^3x\, (s-g\tilde{A})\epsilon \partial A  - K_B[\tilde{A}]} \;,
\end{eqnarray}
and using the bosonization formula (\ref{bos}), this is precisely the Wilson
loop in the noncompact model including the dynamical fermions,
\begin{equation}
W(s)=\frac{1}{{\mathcal N}}\int {\mathcal D}A\, {\mathcal D}\bar{\psi}\, {\mathcal D}\psi \, e^{-\int d^3x\,
\frac{1}{2}f^2+i\int d^3x\, (sf-gA J)  - K_F}\;.
\end{equation}

Then, we clearly see that the effect of the dynamical matter, via the
associated local $Z_2$ symmetry, is that of gauging out the field $\phi$ or,
in an equivalent manner, completely erasing the effect of the instantons. That
is, the theory becomes noncompact.  In the next section, we will see that the
complete erasure of instantons only occurs when the dynamical matter includes
massless fermions, and certain conditions are satisfied, which settle the applicability of
the local $Z_2$ scenario.  In addition, the following remarks
are in order:
\begin{enumerate}
\item[i)]The resulting effect of the local $Z_2$ symmetry is similar to that
obtained when introducing parity-breaking matter, although the physical reason
is entirely different. In the parity-breaking case, the instantons are neatly
erased because of topological reasons associated with the induced Chern-Simons
term coming from the fermionic effective action (see~\cite{fs}). Indeed,
demanding gauge invariance in the presence of instantons, one discovers that
there can be no instanton.
\item[ii)]In the bosonization approach, the $Z_2$ local symmetry of the dual
  model is present even in the large mass limit. As a consequence, as
  discussed in \cite{lucho}, the relevant source for the Wilson loop is still
  concentrated on the boundary, and the expected area law is not re-obtained
  when the large mass limit is taken but a destabilization of the perimeter
  law.
\item[iii)]Of course, the global $Z_2$ sine-Gordon model correctly describes
  the decoupling of large mass fermions and the corresponding confinement of
  the electric test charges associated with the Wilson loop.
\item[iv)]On the other hand, according to our discussion in
  subsection~\ref{ssec:mg}, when the effect of massless fermions is studied by
  means of the anomalous sine-Gordon model with global $Z_2$, given
  in~\cite{kleinert,igor}, a destabilization of the area law is obtained,
  instead of the perimeter law characterizing deconfinement.
\end{enumerate}

\section{Zero modes and the induced instanton 
  anti-instanton interaction}\label{zm} From the discussion in
section~\ref{sec:cwilson}, the question suggests itself about how the fermion
mass singles out the proper low energy description of the system.

In this regard note that for large masses, $\mathcal{K}\sim {\rm const.}$, and
the instanton interaction read from (\ref{ii}) (of course, disregarding the
Wilson loop source $s_\mu$) becomes the usual Coulomb potential in three
dimensions. Then, a dilute monopole plasma calculation is indeed justified,
and this leads to the dual model represented by (\ref{dp}), which displays a
global $Z_2$ symmetry.  Then, the relevant source to compute the Wilson loop
is concentrated on the surface bounded by the loop; this together with the
localization properties of the usual sine-Gordon model ($\mathcal{K}\sim {\rm
  const.}$) lead to the area law.

On the other hand, for massless fermions, as discussed in~\cite{kleinert}, if
the truncation of the fermionic effective action is accepted, according to
(\ref{ii}) with $\mathcal{K}\sim \frac{1}{(\partial^2)^{1/2}}$ the instanton
interaction would be logarithmic. This posses some reservations regarding the
applicability of the dilute monopole plasma approach and the consideration of
the anomalous Sine-Gordon model to compute the Wilson loop. This explains the
destabilization of the area law discussed in section~\ref{sec:cwilson},
instead of the expected perimeter law, when considering this model.

Then, to construct a reliable description of compact $QED_3$ with massless
fermions, it is important to have a good understanding of the effective
instanton interactions in this case. In particular, we should know how the
energy~\footnote{We follow the usual practice of calling energy to the
  quantity which in fact is an action. It may of course be regarded as an
  energy if one thinks of the Euclidean theory in $d$ spacetime dimensions as
  a static theory in $d+1$ dimensions.} of an instanton anti-instanton pair
depends on their distance.

A concrete way to study that effective interaction between an instanton and an
anti-instanton separated by a (spacetime) distance $L$, is to consider the
function $\Omega (L)$
\begin{equation}
\Omega (L) \;\equiv \; -\, 
\ln \Big[ \int {\mathcal D}\psi {\mathcal D}{\bar\psi} 
\, e^{- S_F({\bar\psi}, \psi; A^L)} \Big]
\end{equation}
where $A^L$ denotes the gauge field corresponding to the instanton
anti-instanton pair, and its explicit form is of course dependent on the
number of dimensions. $S_F$ is the Dirac action for fields minimally coupled
to $A^L$. 

In what follows, we present the discussion of the two and three-dimensional
cases.
\subsection{Two dimensions}
In two spacetime dimensions, we have for the $A_L$ gauge field the following
condition: 
\begin{equation}
\epsilon_{\mu\nu} \partial_\mu A^L_\nu (x) \;=\; 
2 \pi \, q \Big[ \delta^{(2)} (x - u)
- \delta^{(2)} (x - v)\Big]
\end{equation} 
where $q$ is the (integer) instanton charge, while $u$ and $v$ are the
coordinate vectors corresponding to the instanton and anti-instanton locii,
respectively.

We then introduce bosonization, in order to calculate $\Omega$. Under this
transformation, the fermionic current $j_\mu$ is mapped into the curl of a
pseudoscalar field $\phi$: $j_\mu \to
\frac{1}{\sqrt{\pi}}\epsilon_{\mu\nu}\partial_\nu \phi$, while the fermionic
action has a different form depending on whether the fermion mass is equal or
different from zero. In the former case, we simply have a quadratic action for
$\phi$:
\begin{equation}\label{eq:quadphi}
S_F \;=\; \frac{1}{2} \, \int d^2x \, ( \partial\phi)^2  \;.
\end{equation}
Thus, for the massless fermion case,
\begin{equation}
\Omega (L) \;=\; - \ln \langle \, e^{i \int d^2x j_\mu  A^L_\mu )}
\rangle \;=\; - \ln \langle \, e^{- \, i \, \sqrt{4 \pi} \, q \,
[\phi(u) - \phi(v) ]} \rangle
\;,
\end{equation}
where $\langle \ldots \rangle$ denotes functional average in the bosonic
theory, with the quadratic action (\ref{eq:quadphi}).  A straightforward
calculation yields:
\begin{equation}
\Omega (L) \;=\; 2 \, q^2 \; \ln | u - v| \;,
\end{equation}
where we have subtracted a contribution corresponding to the self-interactions
(which is, of course, $L$-independent). In this regard, we note that the would
be self-action of the vortex field due to the fermions, is divergent in the
continuum (of course, it will be finite if the vortices are given a finite
core size):
\begin{equation}
\ln \langle e^{- \, i \, \sqrt{4 \pi} \, q \, \phi(x)} \rangle 
\;\to \; \infty  \;\;,\;\;\;\; m = 0\;.
\end{equation} 

In the massive fermion case, the bosonic theory is also exactly known, it
becomes a massive sine-Gordon action:
\begin{equation}\label{eq:sinephi}
S_F \;\to\; \frac{1}{2} \,\int d^2x \big[ (\partial\phi)^2 
+ \Lambda \, \cos( \sqrt{4\pi} \phi) \big] \;, 
\end{equation}
where $\Lambda$ is a mass parameter.

Of course, we still have:
\begin{equation}
\Omega (L) \;=\; -  \ln \langle \, e^{- \, i \,\sqrt{4\pi} \, q\, 
[ \phi(u) - \phi(v) ]} \rangle \;,
\end{equation}
where now $\langle \ldots \rangle$ is evaluated with the action
(\ref{eq:sinephi}). This kind of correlation function is exactly known.
Indeed, introducing the dimensionless variable: $x \equiv m L$, and defining:
\begin{equation}
w_q (x) \;=\; - \Big[ L \frac{d}{dL} \, \Omega (L) \Big]_{x \equiv m L} \;,
\end{equation}
it can be shown that $w_q$ may be expressed in terms of another function $v_q$
\begin{equation}
w_q(x) \;=\; - \int_x^\infty dy \,y \, v_q^2 (y)
\end{equation}
which satisfies the non-linear differential equation:
\begin{equation}
v_q'' + \frac{1}{x} v_q' \;=\; - \frac{v_q}{1 - v_q^2} \, (v_q')^2 + v_q -
v_q^3 + \frac{4 q^2}{x^2} \, \frac{v_q}{1 - v_q^2} \;.
\end{equation}

Rather than giving approximate numerical solutions for the equation above, we
comment on the exact results that may be obtained in some particular cases. We
first note that, when the fermions are massive, the average for a single
vortex is no longer zero~\cite{lukyanov}:
\begin{equation}
\langle e^{- \, i \, \sqrt{4 \pi} \, q \, \phi(x)} \rangle \;=\; 
(\frac{m}{2})^{q^2} \; \exp\left\{ \int_0^\infty \frac{dt}{t} \, \Big[ 
\frac{\sinh^2( q\, t )}{\sinh^2 (t)} - q^2 e^{- 2 t} \Big]\right\} 
\;\;.
\end{equation}
This means that the would-be vortex self-action $E \equiv -\ln\langle e^{-
  i\sqrt{4 \pi} q \phi}\rangle$, goes like $E \sim -\ln (a m)$, where $a$
denotes a small length scale, associated to the vortex size. This length scale
defines the mass $m_0$ for which the self-action of the vortex is zero, which
turns out to be $m_0 \sim 1/a$, i.e., of the order of the cutoff. This is to
be expected, since no quantum effect would come from those heavy states.

Besides, one can also show~\cite{lukyanov} that 
\begin{equation}
\langle \, e^{- \, i \,\sqrt{4\pi} \, q\, [ \phi(u) - \phi(v) ]} \rangle
\;\to\; \langle e^{- i\sqrt{4 \pi} q \phi}\rangle
\langle e^{i\sqrt{4 \pi} q \phi}\rangle \;\;,\;\;\;\; L \to \infty \;.
\end{equation}
In the massless case, we know that each factor is zero, and that the
correlation function tends to zero as a power.  In the massive case, however,
we see that the vortex ($q=1$) antivortex ($q=-1$) interaction energy coming
from the fermionic sector tends to a finite constant when they are separated
by an infinite distance:
\begin{equation}
\Omega (L) \;\sim \; - 2 \,\ln (a m)  \;\;,\;\;\; 
m \neq 0\;,\;\;\; L \to \infty \;. 
\end{equation} 
We conclude that, in the massive case, the additional vortex interaction
coming from the fermion sector is suppressed with respect to the logarithmic
vortex interaction coming from the gauge field fluctuations. In other words,
the effect of massive fermions is irrelevant. In the massless case, the
induced vortex interaction is logarithmic thus renormalizing the effect of the
gauge field fluctuations.

\subsection{Three dimensions}

In the $2+1$ dimensional case, we have already seen that the effect of large
mass fermions is inducing a fermionic effective action of the (local) Maxwell
form, simply renormalizing the instanton $1/L$ Coulomb interaction associated
with the gauge field fluctuations.  In this case, when an
instanton/anti-instanton pair is far apart, the fermionic effective action
tends to a constant, or in other words, the fermion determinant does not
suppress the instanton contributions.

On the other hand, for massless fermions we do not have the exact solutions of
the $1+1$ dimensional case at our disposal. However, by using an approximated
analytical treatment, which assumes that the monopole and antimonopole are far
apart, one can find the effective interaction, even at finite
temperature~\cite{Agasian:2001an} in the Georgi-Glashow model. This model is
of course known to be a `regularized' form of compact $QED_3$. The $T=0$
result is found to be~\cite{Agasian:2001an}:
\begin{equation}
\Omega(L) \;=\;  \ln ( m^2 + |a|^2 )
\end{equation}
where $a$ is a matrix element obtained from the overlap between the zero modes
corresponding to fermions in the background of a monopole and an antimonopole
(with different centers).  Explicitly,
\begin{equation}
|a| \;=\; - 4 \pi \, \ln ( \mu L) \;,
\end{equation}
where $\mu$ is an IR cutoff.

We note that in the $m\to 0$ limit, at $T=0$ and for large $L$, the induced
interaction coming from the massless fermion sector is not logarithmic,
implying that the fermionic determinant is not suppressed as a power of $1/L$.
In other words, the fermionic effective action in (\ref{nl}) does not capture
the instanton interaction at $T=0$, as it would imply a logarithmic one.
However, according to the discussion in~\cite{Agasian:2001an} we have the
following situation:
\begin{enumerate}
\item[i)]At high temperatures, dimensional reduction occurs. In the pure
  compact theory, the instanton flux lines are spread essentially on a 2D
  region and this gives logarithmic interactions between instantons, which can
  be associated with a BKT transition. Above the critical temperature $T_c$
  the instantons become suppressed forming dipoles. This corresponds to
  deconfinement at high temperatures.
\item[ii)]When massless fermions are included, the critical temperature $T_c$,
  above which the mean instanton distance is finite, is drastically lowered so
  that the dipole phase survives up to very low temperatures.
\item[iii)] In this phase, for large distances, the fermionic determinant is
  suppressed as an inverse power of $L$: $1/L^{2N_f}$. This means that above
  $T_c$ the anomalous quadratic contribution in (\ref{nl}) captures the
  logarithmic behavior of the instanton interactions.  As $T_c$ is
  exponentially suppressed, this means that the anomalous model can be used up
  to very low temperatures.
\item[iv)]In the London limit, where the mass of the $W_\pm$ modes tends to
  infinity, the mean instanton distance in the molecules collapses to zero.
  Then, in the London limit, which corresponds to compact $QED_3$, the theory
  including massless fermions becomes essentially noncompact.
\end{enumerate}
Because of i)-iv), the phase associated with compact $QED_3$ when including
massless fermions, can be studied in terms of the noncompact version of the
model.  Taking into account the anomalous induced action in (\ref{nl}), and
according to the detailed calculation in subsection~\ref{ssec:qed3}, this
means a $1/R$ interaction between static charges and a Wilson loop perimeter
law characterizing a deconfining phase.

Note also that, as discussed in subsection~\ref{boscom}, the noncompact model
can be also rephrased in terms of a local $Z_2$ symmetry.

\section{Instanton dipole liquid model: stability of the perimeter law} \label{idl}

The considerations above justify an effective dual model to describe the
instanton phase above $T_c$, taking as starting point the noncompact theory
including the anomalous quadratic action in (\ref{nl}). This anomalous
noncompact model represents the situation where the instantons are completely
erased. Thus representing the $N_f\geq 2$ London limit of the Georgi-Glashow
model. Deviations from this regime, as a large but finite mass for the $W_\pm$
modes or a small but nonzero fermion mass, can be analyzed by incorporating
the formation of dipole configurations.

We have already seen that the summation over instantons cannot be implemented
as an anomalous instanton gas. We propose instead the summation over
configurations of the instanton dipole liquid type.  That is,
\begin{equation}
W(s)=\int {\cal D}M \; e^{- S(M)}  \int {\cal D}A\; e^{- \int d^3x\,\big( 
\frac{1}{2}f_\mu {\mathcal O} f_\mu + i  A_\mu m_\mu \,-\, 
i  s_\mu f_\mu \big)}\; ,
\label{pro}
\end{equation}
where $M_\mu$ is the magnetization field associated with the dipole instanton
liquid and $m\equiv \epsilon \partial M$ is the corresponding (topologically
conserved) instanton current. We have included a minimal coupling between that
conserved current and the gauge field (the $i$ is required in Euclidean
spacetime), and $S(M)$ denotes the action for the magnetization field, about
which we shall say something below.

The kernel ${\mathcal O}$ contains both the initial (pure) Maxwell term plus
the nonlocal induced term in (\ref{nl}),
\begin{equation}
{\mathcal O} \;=\; 1 \,+\, \xi \frac{g^2}{\sqrt{-\partial^2}} \;,
\end{equation} 
note that, for large distances, the second term dominates over the local
Maxwell term.

For the sake of simplicity, we will assume that $S(M)$ is a quadratic
functional of the magnetization,
\begin{equation}
S(M) \;=\; \frac{1}{2}\, \int d^3x \, M_\mu \Omega M_\mu\; ,
\end{equation}
where $\Omega$ is the kernel that defines the action for the magnetization
field $M$. By symmetry reasons, it is reasonable to assume it to be a scalar
function of the Laplacian.
 
Of course, the integral over $M$ and $A$ can be performed exactly with the
outcome:
\begin{equation}
W(s)\;=\; 
e^{- \frac{1}{2} \, \int d^3x\; s^\perp_\mu \big ( {\mathcal O} \,+\,
  \Omega^{-1}\big)^{-1}  s_\mu^\perp } \;,
\end{equation}
where $s^\perp$ is the transverse part of the current localized on the
surface of the loop.
 
Now, let us first assume the simplest (local) form for $\Omega$: $\Omega =
1/a^2$, where $a$ is a real constant.  This represents a system where large
magnetizations are suppressed, so that it describes a basic feature of the
phase where the instanton dipole liquid is formed.  In particular, note that
in the $a\to 0$ limit, the magnetization is completely eliminated and the
noncompact situation is re-obtained. In the general case, when a finite $a$ is
considered, we see that:
\begin{equation}
 \ln \Big[ \langle W({\mathcal C}) \rangle \Big]_{reg}  \;\equiv\;
 - \, \frac{1}{2} \, g^2 \, \mu^{1 - \alpha} \; \int \frac{d^3k}{(2\pi)^3} \,
  \frac{|{\mathbf k}|^{1 + \alpha}}{k^2} \; 
\frac{1}{1  + \xi \frac{g^2}{k} + a^2} \, 
|{\tilde \chi}_\Sigma ({\mathbf k}) |^2 \;,
\end{equation}
where we adopted the same regularization as in the noncompact case.

This may be easily rearranged to look identical to the analog noncompact
expression:
\begin{equation}
 \ln \Big[ \langle W({\mathcal C}) \rangle \Big]_{reg}  \;\equiv\;
 - \, \frac{1}{2} \, g_a^2 \, \mu^{1 - \alpha} \; \int \frac{d^3k}{(2\pi)^3} \,
  \frac{|{\mathbf k}|^{1 + \alpha}}{k^2} \; 
\frac{1}{1  + \xi \frac{g_a^2}{k}} \, 
|{\tilde \chi}_\Sigma ({\mathbf k}) |^2 \;,
\end{equation}
with $g_a \equiv g/\sqrt{ 1 + a^2}$. Therefore, in this case, there is no
departure from the perimeter law obtained in the noncompact case, except for a
finite renormalization of the coupling constant.

This comes about since the object that determines the behaviour of the Wilson loop average is $({\mathcal O}\,+\, \Omega^{-1}\big)^{-1}$, so it is clear that the IR (long distance) behaviour of the loop might only be changed, with respect to the one dictated by ${\mathcal O}$, if $\Omega^{-1}$ produced a stronger divergence for small momenta. Otherwise,
there are not relevant changes. 

For this reason, if the above mentioned model is improved by including correlations
in the magnetization at different points, $\Omega=\frac{1}{\kappa^2} \, (-\partial^2) +1/a^2$, as the infrared behavior of ${\mathcal O}$ continues to dominate over the effect of the magnetization, this model will also display a perimeter law.  

An example where the IR behaviour may be changed corresponds to the $a \to
\infty$ limit in the above expression for $\Omega$, that is, $\Omega =
\frac{1}{\kappa^2} \, (-\partial^2)$. Considering the usual circular region in
the $x_3$ plane, this leads, in Fourier space, to the following expression for
the Wilson loop average:
\begin{equation}
\ln \Big[\langle W({\mathcal C}) \rangle \Big] \;=\;
- \frac{1}{2} \,\int \, \frac{d^3k}{(2\pi)^3} \; \frac{{\mathbf k}^2}{k^2} \;
| {\tilde \chi}_\Sigma ({\mathbf k}) |^2 \;
\frac{1}{1 + \xi \frac{g^2}{k} + \frac{\kappa^2}{k^2}}
\;,
\end{equation}
where ${\tilde \chi}_\Sigma$ is the (two dimensional) Fourier transform of the
characteristic function for the circle. Using then the explicit form for the
kernels ${\mathcal O}$ and $\Omega$, we find:
\begin{equation}
\ln \Big[\langle W({\mathcal C}) \rangle \Big] \;=\;
- \frac{1}{2} \,R^2\, \int_{-\infty}^{+\infty} \, dk_0  
\; \int_0^\infty d|{\mathbf
  k}| \, |{\mathbf k}| \; \frac{J_1^2(|{\mathbf k}| R)}{k^2 + \xi g^2 k 
+ \kappa^2} \;,
\end{equation}
which is UV divergent. Including the same kind of analytic regularization we
applied in the free case, we see that: 
\begin{equation}
\ln \Big[\langle W({\mathcal C}) \rangle \Big]_{reg}
\;=\;
- \frac{1}{2} \,R^2\, (\mu)^{1-\alpha} \,\int_{-\infty}^{+\infty} \, dk_0  
\; \int_0^\infty d|{\mathbf
  k}| \, |{\mathbf k}|^{\alpha} \; \frac{J_1^2(|{\mathbf k}| R)}{k^2 + 
\xi g^2 |k| + \kappa^2} \;,
\end{equation}
or, by a change of variables,
\begin{eqnarray}
\ln \Big[\langle W({\mathcal C}) \rangle \Big]_{reg}
&=&
- \frac{1}{2} \,R \, (\mu R)^{1-\alpha} \,
\times \int_{-\infty}^{+\infty} \, dl_0  
\; \int_0^\infty d|{\mathbf l}| \nonumber\\ 
&\times & |{\mathbf l}|^{\alpha} \; \frac{J_1^2(|{\mathbf l}|)}{l^2 + 
\xi g^2 |l| R + \kappa^2 R^2} \;.
\end{eqnarray}
It should be evident (from the discussion in the free case) that the integral
over $l_\mu$ shall produce a pole in $\varepsilon = 1-\alpha$, and hence the
result, for long distances, will be essentially $R \ln R$.

Of course, the destabilization of the perimeter law is expected in this case,
as an action for the magnetization based on the operator $\Omega =
\frac{1}{\kappa^2} \, (-\partial^2)$ would also represent a destabilization of
the instanton dipole liquid: for homogeneous configurations there would be no
cost to increase the magnetization, or equivalently, the instanton
anti-instanton distance.

\section*{Conclusions}
In this article we have studied the mechanism by which the mass of (parity-
preserving) fermions singles out the proper low-energy dual description of
compact $QED_3$, explaining whether the resulting phases are confining or
deconfining. To that end, we have discussed the instanton anti-instanton
interactions.

For a large fermion mass, the quadratic approximation to the fermionic
effective action was found to be reliable, giving a local Maxwell term which
renormalizes the pure-gauge action. In this case, the instanton anti-instanton
interaction obeys the usual Coulomb behavior.  Then, a dilute monopole gas
approximation can be considered to implement the compactification of the
model, and the usual Polyakov dual model (with renormalized parameters) is
obtained.  This model possesses a global $Z_2$ symmetry, so that the relevant
source to compute the Wilson loop is concentrated on the surface bounded by
the loop.  Therefore, the area law characterizing confinement is obtained in
this case.

On the other hand, it is known that when two or more flavours of massless
fermions are considered~\footnote{Note that changing the number of flavours
  requires, in our notation, to change the value of the constant $\xi$.},
above an exponentially suppressed critical temperature $T_c$, the instantons
bind to anti-instantons to form dipoles.

In the Georgi-Glashow regularized version of compact $QED_3$, the average
instanton anti-instanton distance in the molecule is finite; however, in the
London limit where compact $QED_3$ is approached, the average distance
collapses to zero.  Then, in this situation, the effect of the massless
fermions is that of completely erasing the instanton configurations, thus
leading to an effectively noncompact gauge model. This model can be
characterized by the particular suppression of the fermionic determinant, due
to quasi zero modes, when the instanton and anti-instanton are separated by a
large distance $L$.  As this suppression goes like an inverse power of $L$, it
amounts to an attractive logarithmic instanton interaction, which is captured
by the behavior arising from the quadratic truncation of the fermionic
determinant.

Summarizing, in the above mentioned conditions, the presence of massless
fermions in compact $QED_3$ render the theory essentially noncompact, and
hence the anomalous effective action induced by the massless fermions is a
good starting point to study the corresponding phase. We have presented a
careful calculation of the Wilson loop in this model to show that the
perimeter law is indeed satisfied, and the corresponding phase is deconfining.
This outcome is similar to that observed in compact $QED_3$ with parity
breaking matter where the instantons were also erased, but due to topological
reasons associated with the induced Chern-Simons term.

On the other hand, for a large mass of the $W_\pm$ modes in the Georgi-Glashow
model, but outside the London limit, or for nonzero but very small fermion
masses, the average dipole size is expected to be finite. For this reason, we
have also studied deviations from the noncompact situation by considering a
model for the instanton dipole liquid formed in this case, showing that the
perimeter law is preserved for a reasonable class of actions characterizing
the magnetization field for a liquid of stable instanton dipoles.

Finally, with regard to the symmetry properties characterizing the low energy
dual description of $QED_3$, we have also shown, via bosonization, that the
erasing of instantons can be associated to a situation where the initial
global $Z_2$ symmetry in the pure gauge theory becomes local when massless
fermions are included.  This is to be compared with an anomalous dual model
for compact $QED_3$ with massless fermions, previously discussed in the
literature, possessing a global $Z_2$ symmetry.  Here, we have shown that such
a global $Z_2$ scenario would give a destabilization of the area law, instead
of the expected perimeter law associated with deconfinement.

Then, by tuning the fermion mass, from very large to very small values, there
is a confinement/deconfinement transition which cannot be described by means
of a single dual low energy effective model. The associated low energy
descriptions are essentially different, displaying a global or local $Z_2$
symmetry, respectively.

\section*{Acknowledgements}
C.~D.~Fosco is a member of CONICET (Argentina).  The SeCTyP-U.N.Cuyo (Argentina) (C.D.F.), the Conselho Nacional de Desenvolvimento Cient\'{\i}fico e Tecnol\'ogico (CNPq-Brazil) (L.E.O.) and the Funda\c c\~ao de Amparo \`a Pesquisa do Estado do Rio de Janeiro (FAPERJ-Brazil) (L.E.O.) are acknowledged for the financial support.


\end{document}